\documentclass[12pt,a4paper]{article}
\usepackage{latexsym,amsthm,amssymb}
\newtheorem{theorem}{Theorem}
\newtheorem{proposition}[theorem]{Proposition}

\newtheorem{corollary}[theorem]{Corollary}
\newtheorem{lemma}[theorem]{Lemma}
\newcommand{\be}{\begin{equation}}
\newcommand{\ee}{\end{equation}}
\newcommand{\bea}{\begin{eqnarray}}
\newcommand{\eea}{\end{eqnarray}}
\newcommand{\ba}{\begin{array}}
\newcommand{\ea}{\end{array}}
\newcommand{\bean}{\begin{eqnarray*}}
\newcommand{\eean}{\end{eqnarray*}}
\newcommand{\no}{\nonumber}

\newcommand{\Ga}{\Gamma}

\newcommand{\la}{\lambda}

\newcommand{\de}{\delta}

\newcommand{\pa}{\partial}

\newcommand{\res}{{\rm res\/}}

\newcommand{\ep}{\epsilon}

\newcommand{\taub}{\tau}

\begin{document}
\title
     {\bf  On the BKP hierarchy: additional symmetries, Fay identity and Adler-Shiota-van
     Moerbeke formula\/}
\author
{\sc Ming-Hsien Tu\footnote{E-mail: phymhtu@ccu.edu.tw} \\ \\
    {\it Department of Physics, National Chung Cheng University\/},\\
   {\it Minghsiung, Chiayi 621, Taiwan\/}\/}
\date{\today}
\maketitle
  \begin{abstract}
We give an alternative proof of the Adler-Shiota-van Moerbeke formula for
the BKP hierarchy. The proof is based on a simple expression for the generator of
additional symmetries and the Fay identity of the BKP hierarchy.
  \\
  \\
PACS: 02.30.Ik\\
Mathematics Subject Classification (2000): 35Q58, 37K10\\
Keywords:  BKP hierarchy, vertex operators, additional symmetries, Fay identity,
Adler-Shiota-van Moerbeke formula.
\end{abstract}
\newpage
%%%%%%%%%%%%%%%%%%%%%%%%%%%%%%%%%%%%%%%%%%%%
\section{Introduction}
%%%%%%%%%%%%%%%%%%%%%%%%%%%%%%%%%%%%%%%%%%%%%
There are two important symmetries associated with the Kadomtsev-Petviashvili (KP)
hierarchy\cite{DJKM}. One is the Sato's B\"acklund symmetry generated by vertex operator
\cite{DJKM} and the other the additional symmetries by Orlov-Schulman operator\cite{OS86,Dic93}.
They have been realized  from such several points of
view as conformal algebras and string equation of matrix models in the 2-d
quantum gravity  etc.(see e.g. \cite{vM} and references therein)
Both of them commute with the hierarchy flows but do not
commute between themselves. These two symmetries looking quite different initially  are in fact
related to each other via the so-called Adler-Shiota-van Moerbeke (ASV) formula \cite{ASM94,ASM95}
which states that additional symmetries acting on the wave function is equivalent to
Sato's B\"acklund symmetries acting on the corresponding tau function, and
the associated algebra of additional symmetries is then lifted to its central extension,
the algebra of B\"acklund symmetries.
The proof was then simplified by Dickey \cite{Dic95b} using the notion of resolvent operators
 and the Fay identity of the KP hierarchy.

In this work, we consider the KP hierarchy of $B$ type \cite{DJKM}
(or BKP hierarchy for short) which is a reduction
of the ordinary KP hierarchy (or KP hierarchy of $A$ type). Here $B$ stands for the infinite
dimensional Lie algebra $go(\infty)$, in contrast to $gl(\infty)$ for the KP.
The BKP hierarchy possesses many integrable structures as the KP hierarchy, for example,
Lax equations, linear system, $\tau$-function, Hirota bilinear form, free fermion representation,
soliton  and quasi-periodic solutions etc. (see \cite{DJKM} for a review)
In \cite{L95} van de Leur obtained the corresponding ASV formula for the BKP hierarchy,
which is similar to but different from that of the
KP hierarchy. Motivated by the KP hierarchy, we attempt to give an alternative proof of
the ASV formula for the BKP hierarchy based on Dickey's approach\cite{Dic95a,Dic95b}.
Our main results contain two parts. The first part is to give a simple expression of the generator
of additional symmetries of the BKP hierarchy, which as a by-product provides an origin of the
special form of the Lax operator for the constrained BKP(cBKP) hierarchy\cite{YC92,LW99}.
The second part is to derive the Fay identity and its differential form from the
bilinear identity which together with the first part gives a conceptually  simple proof
of the ASV formula for the BKP hierarchy. Let us recall some basic properties of the BKP hierarchy.

%%%%%%%%%%%%%%%%%%%%
\section{The BKP hierarchy}
%%%%%%%%%%%%%%%%%%%%
The BKP hierarchy \cite{DJKM} can be defined by the Lax equation
\be
\pa_{2n+1}L=[B_{2n+1},L],\quad B_{2n+1}=(L^{2n+1})_+,\quad n=0,1,2,\cdots
\label{laxeq}
\ee
where the Lax operator has the form
\[
L=\pa+u_1\pa^{-1}+u_2\pa^{-2}+\cdots,
\label{laxop}
\]
with coefficient functions $u_i$ depending on the time variables $t=(t_1=x,t_3,t_5,\cdots)$
and satisfies the constraint
\be
L^*=-\pa L\pa^{-1}.
\label{constr}
\ee
Here we use the notations: $\pa_{2n+1}f=\pa f/\pa t_{2n+1}$,
$(\sum_ia_i\pa^i)_+=\sum_{i\geq 0}a_i\pa^i$,
$(\sum_ia_i\pa^i)_-=\sum_{i< 0}a_i\pa^i$, $(\sum_ia_i\pa^i)_{[k]}=a_k$ and
$(\sum_ia_i\pa^i)^*=\sum_i(-\pa)^ia_i$.
It can be shown \cite{DJKM} that the constraint (\ref{constr}) is
equivalent to the condition $(B_{2n+1})_{[0]}=0$.

The Lax equation (\ref{laxeq}) can be described by the compatibility condition of the
linear system
\be
Lw(t,\la)=\la w(t,\la),\qquad \pa_{2n+1}w(t,\la)=B_{2n+1}w(t,\la),
\label{lineq}
\ee
where $w(t,z)$ is called wave function (or Baker function) of the system and $\la$
is the spectral parameter.
The whole hierarchy can be expressed in terms of a dressing operator, the so-called
Sato operator $W$\cite{DJKM}, as
\be
L=W\pa W^{-1},\qquad w(t,\la)=We^{\xi(t,\la)},
\label{dresslax}
\ee
where $\xi(t,\la)=\sum_{i=0}t_{2i+1}\la^{2i+1}$ and $W=1+\sum_{i=1}^\infty w_i\pa^{-i}$.
Then the Lax equation (\ref{laxeq}) is equivalent to the Sato equation
\be
\pa_{2n+1}W=-(L^{2n+1})_-W
\label{satoeq}
\ee
where we refer $(L^{2n+1})_-$ to the generator of inner symmetries of the hierarchy.
Using the dressing form (\ref{dresslax}), the constraint (\ref{constr}) can be
expressed as\cite{S89,T93}
\be
W^*\pa W=\pa.
\label{constrsato}
\ee
From (\ref{dresslax}) the solutions of the linear system (\ref{lineq}) has the form
\be
w(t,\la)=\hat{w}(t,\la)e^{\xi(t,\la)},
\label{wavefun}
\ee
where $\hat{w}(t,\la)=1+w_1/\la+w_2/\la^2+\cdots$.
\begin{lemma}\label{PQ}
 For pseudodifferential operators $P,Q$, the following formula holds.
\[
\res_\la[\la^{-1}(\pa^jPe^{\la x})(Qe^{-\la x})]
=\res_\pa(\pa^jP\pa^{-1}Q^*), \quad j\geq 0,
\]
where we denote the symbols $\res_z(\sum_ia_iz^i)=a_{-1}$ and $\res_\pa(\sum_ib_i\pa^i)=b_{-1}$.
\end{lemma}
Using Lemma \ref{PQ} it can be shown \cite{DJKM} that $w(t,\la)$ is a wave function of
the BKP hierarchy if and only if it satisfies the bilinear identity
\be
\res_\la(\la^{-1}w(t,\la)w(t',-\la))=1.
\label{bileq}
\ee
In fact, from the bilinear identity (\ref{bileq}), solutions of the BKP hierarchy
can be characterized by a single function $\tau(t)$ called $\tau$-function
such that\cite{DJKM}
\be
\hat{w}(t,\la)=\frac{\tau
(t_1-\frac{2}{\la},t_3-\frac{2}{3\la^3},t_5-\frac{2}{5\la^5},\cdots)}
{\tau(t)}\equiv\frac{\tau(t-2[\la^{-1}])}{\tau(t)}.
\label{tau}
\ee
%%%%%%%%%%%%%%%%%%%%%%%%%%%%%%%%%%%%%%%%%%%%%%%%%%%%%%%%%%%%%%
\section{Vertex operators and Sato's B\"acklund transformations}
%%%%%%%%%%%%%%%%%%%%%%%%%%%%%%%%%%%%%%%%%%%%%%%%%%%%%%%%%%
From (\ref{wavefun}) and (\ref{tau}) the wave function $w(t,\la)$ can be written as
\[
w(t,\la)=\frac{X_B(t,\la)\taub(t)}{\taub(t)},
\]
where $X_B(t,\la)$ is the so-called vertex operator, defined by \cite{DJKM}
\[
X_B(t,\la)=e^{\xi(t,\la)}e^{-2\sum_{n=0}^\infty\frac{\la^{-2n-1}}{2n+1}\pa_{2n+1}}
\equiv e^{\xi(t,\la)}G(\la).
\]
Also another useful vertex operator $X_B(\la,\mu)$ can be defined as
\[
X_B(\la,\mu)=e^{-\xi(t,\la)}e^{\xi(t,\mu)}G(-\la)G(\mu).
\]
\begin{proposition}(Sato's B\"acklund transformations\cite{DJKM})
The vertex operator $X_B(\la,\mu)$
provides  infinitesimal transformations on the space of $\tau$-function, namely, if
$\tau(t)$ is a solution then $\tau(t)+\ep X_B(\la,\mu)\tau(t)$ is a solution as well.
\end{proposition}
Introducing the symbol $\alpha(z)=\sum_{n\in{\mathbb Z\/}_{odd}}\alpha_nz^{-n}/n$ with
\[
\alpha_n=
\left\{
\ba{ll}
2\pa/\pa t_n & n>0 \\
|n|t_{|n|}& n<0
\ea
\right.
\]
that satisfy the commutation relation
\[
[\alpha_n,\alpha_m]=2n\de_{n,-m},\quad n\in {\mathbb Z\/}_{odd}.
\]
Then the vertex operator $X_B(\la,\mu)$ can be expressed as
\[
X_B(\la,\mu)=:e^{\alpha(\la)-\alpha(\mu)}:
\]
where the normal ordering $::$ demands that $\alpha_{n>0}$ must be placed to the
right of $\alpha_{n<0}$. One can Taylor expand $X_B(\la,\mu)$
around $\mu=\la$ for large $\la$ as
\[
X_B(\la,\mu)=\sum_{m=0}\frac{(\mu-\la)^m}{m!}W^{(m)}(\la)
=\sum_{m=0}\frac{(\mu-\la)^m}{m!}\sum_{l=-\infty}^\infty\la^{-m-l}W_l^{(m)},
\]
with
\[
W^{(m)}(\la)=\pa_\mu^mX_B(\la,\mu)|_{\mu=\la}=:\pa_\la^me^{-\alpha(\la)}\cdot e^{\alpha(\la)}:
=:P_m(-\pa_\la\alpha(\la)):
\]
where $P_m(u(z))$ is the so-called Fa\`a di Bruno polynomials (see e.g. \cite{Dic03}) defined by
the recurrence relations $P_{m+1}(u)=(\pa_z+u)P_m(u)$. For instance,
$P_0=1, P_1=u, P_2=u'+u^2, P_3=u''+3uu'+u^3$. Then differential operators $W_l^{(m)}$
can be computed as
\bean
W_n^{(0)}&=&\de_{n,0},\\
W_n^{(1)}&=&
\left\{
\ba{ll}
\alpha_n & n\in {\mathbb Z\/}_{odd} \\
0& n\in {\mathbb Z\/}_{even}
\ea
\right.,\\
W_n^{(2)}&=&
\left\{
\ba{ll}
-(n+1)\alpha_n& n\in {\mathbb Z\/}_{odd}  \\
\sum_{i+j=n}:\alpha_i\alpha_j: & n\in {\mathbb Z\/}_{even}
\ea
\right.,\\
W_n^{(3)}&=&
\left\{
\ba{ll}
(n+1)(n+2)\alpha_n-\sum_{i+j+k=n}:\alpha_i\alpha_j\alpha_k:& n\in {\mathbb Z\/}_{odd} \\
-\frac{3}{2}(n+2)\sum_{i+j=n}:\alpha_i\alpha_j: & n\in {\mathbb Z\/}_{even}
\ea
\right.,
\eean
etc.
Moreover, through the fermion-boson correspondence, the representation of Lie algebra
$go(\infty)$ on ${\mathbb C\/}[t_1,t_3,t_5,\cdots]$ is given by\cite{DJKM}
\[
Z_B(\la,\mu)=\frac{1}{2}\frac{\mu+\la}{\mu-\la}(X_B(\la,\mu)-1),
\label{ZB}
\]
which after Taylor expanding around $\mu=\la$ for large $\la$ has the form
\be
Z_B(\la,\mu)=\sum_{m=0}\frac{(\mu-\la)^m}{m!}\sum_{l=-\infty}^\infty\la^{-m-l}Z_l^{(m+1)}.
\label{Zml}
\ee
It is easy to show that differential operators $Z_l^{(m)}$ are related to $W_l^{(m)}$ as
\[
Z_l^{(1)}=W_l^{(1)},\quad
Z_l^{(m+1)}=\frac{W_l^{(m+1)}}{m+1}+\frac{1}{2}W_l^{(m)},\quad m\geq 1,
\]
and constitute  an infinite-dimensional Lie algebra called
 $W^B_{1+\infty}$-algebra which is a subalgebra of $W_{1+\infty}$ associated
 with the KP hierarchy.

%%%%%%%%%%%%%%%%%%%%%%%%%%%%%%%%%%%%%%%%%%%%%%%%%%%%%%%%%%%%%%%%%%
\section{Orlov-Schulman operator and additional symmetries}
%%%%%%%%%%%%%%%%%%%%%%%%%%%%%%%%%%%%%%%%%%%%%%%%%%%%%%%%%%%%%%%%%%
Due to the work of Orlov and Schulman \cite{OS86}, the Lax formulation can be extended by
introducing the Orlov-Schulman operator $M$ defined by
\be
M=W\Ga W^{-1},\quad \Ga=\sum_{n=0}(2n+1)t_{2n+1}\pa^{2n},
\label{dressm}
\ee
which satisfies
\[
\pa_{2n+1}M=[B_{2n+1},M],\quad [L,M]=1.
\]
Thus the linear system (\ref{lineq}) should be extended to
\[
Lw=zw,\quad Mw=\pa_zw,\quad \pa_{2n+1}w=B_{2n+1}w.
\]
Note that on the space of wave function $w(t,z)$, $(L,M)$ is anti-isomorphic to $(z,\pa_z)$
since $[z,\pa_z]=-1$. More general, one has $M^mL^lw=z^l\pa_z^mw,\quad L^lM^mw=\pa_z^mz^lw$,
and $\pa_{2n+1}(M^mL^l)=[B_{2n+1}, M^mL^l]$.
On the other hand, one can introduce the adjoint wave function $w^*(t,z)$ and the adjoint
Orlov-Schulman operator $M^*$ as
\be
w^*(t,z)=(W^*)^{-1}e^{-\xi(t,z)}=-z^{-1}w_x(t,-z),
\label{adjwf}
\ee
and
\be
M^*=(W\Ga W^{-1})^*=(L^*)^{-1}\pa M\pa^{-1}L^*,
\label{constrM}
\ee
where we have used (\ref{constrsato}) and $\Ga^*=\Ga$. Then
$[L^*,M^*]=[M,L]^*=-1$, and
\[
L^*w^*=zw^*,\quad M^*w^*=\pa_zw^*,\quad \pa_{2n+1}w^*=-B_{2n+1}^*w^*.
\]
Now let us introduce a new set of parameters $\hat{t}_{ml}$ to the system
so that additional symmetries of the BKP hierarchy can be expressed as
\be
\hat{\pa}_{ml}W=-(A_{ml}(L,M))_-W,\quad \hat{\pa}_{ml}=\pa/\pa \hat{t}_{ml}
\label{addsato}
\ee
where $A_{ml}(L,M)$ are monomials in $L$ and $M$ and
$A(L,M)=\sum_{ml}c_{ml}A_{ml}(L,M)$ is a generator of additional symmetries.
Then through the dressing formulas (\ref{dresslax}) and (\ref{dressm}) we have
\[
\hat{\pa}_{ml}L=-[(A_{ml}(L,M))_-,L],\quad \hat{\pa}_{ml}M=-[(A_{ml}(L,M))_-,M],
\]
and
\be
\hat{\pa}_{ml}A_{nk}(L,M)=-[(A_{ml}(L,M))_-,A_{nk}(L,M)].
\label{adda}
\ee
\begin{proposition} The additional flows  commute with the hierarchy flows, i.e.
\[
[\hat{\pa}_{ml},\pa_{2k+1}]=0.
\]
Thus they are symmetries of the BKP hierarchy.
\end{proposition}
Proof. From (\ref{satoeq}) and (\ref{addsato}) we have
\bean
[\hat{\pa}_{ml},\pa_{2k+1}]W
&=&-\hat{\pa}_{ml}L^{2k+1}_-W+\pa_{2k+1}(A_{ml}(L,M))_-W,\\
&=&[(A_{ml}(L,M))_-, L^{2k+1}]_-W+L^{2k+1}_-(A_{ml}(L,M))_-W\\
&&+[L^{2k+1}_+,A_{ml}(L,M)]_-W-(A_{ml}(L,M))_-L^{2k+1}_-W,\\
&=&[(A_{ml}(L,M))_-, L^{2k+1}_-]_-W+[L^{2k+1}_-,(A_{ml}(L,M))_-]W=0.\quad \blacksquare
\eean
For the KP hierarchy, $A_{ml}(L,M)=M^mL^l$ and there is no restriction to the
coefficients $c_{ml}$. Due to the anti-isomorphic between $(L,M)$ and $(z,\pa_z)$:
$M^mL^l \mapsto z^l\pa_z^m$ there is a $w_{\infty}$-algebra (or centerless $W_{1+\infty}$-algebra)
associated with additional symmetries of the KP hierarchy, which is generated
by $\{z^{m+l}\pa_z^m, m\geq 0, l\in {\mathbb Z \/}\}$\cite{OS86}.
For the BKP hierarchy the additional flows (\ref{addsato}) should be compatible with
the constraint equation (\ref{constr}).
Therefore one has $\hat{\pa}_{ml}W^*=\pa\cdot \hat{\pa}_{ml}W^{-1}\cdot\pa^{-1}$
which together with (\ref{addsato}) implies that $\pa^{-1}(A_{ml}(L,M)^*)_-\pa=-(A_{ml}(L,M))_-$.
However it is sufficient to consider the condition\cite{T93}
\be
\pa^{-1}A_{ml}(L,M)^*\pa=-A_{ml}(L,M),
\label{addcond}
\ee
which is crucial to determine the algebra associated with additional symmetries.
Substituting $M^mL^l$ into (\ref{addcond}) and using (\ref{constrM}) we have
$\pa^{-1}(M^mL^l)^*\pa=(-1)^lL^{l-1}M^mL$ and thus
the additional flows of the BKP hierarchy are generated by
\[
A_{ml}(L,M)=M^mL^{l}-(-1)^{l}L^{l-1}M^mL.
\]
\begin{proposition} The noncommutative vector fields $\hat{\pa}_{ml}$ acting on
Sato operator  form a $w^B_{\infty}$-algebra (centerless $W^B_{1+\infty}$-algebra):
\[
[\hat{\pa}_{ml},\hat{\pa}_{nk}]=\sum_{pq}C_{nk,ml}^{pq}\hat{\pa}_{pq},
\]
where $C_{nk,ml}^{pq}$ are structure constants of the algebra.
\end{proposition}
Proof. Using (\ref{addsato}) and (\ref{adda}) we have
\[
[\hat{\pa}_{ml},\hat{\pa}_{nk}]W=[A_{ml}(L,M),A_{nk}(L,M)]_-W.
\]
Note that $[A_{ml}(L,M),A_{nk}(L,M)]$  satisfies the constraint (\ref{addcond}), i.e.,
\bean
\pa^{-1}[A_{ml}(L,M),A_{nk}(L,M)]^*\pa=-[A_{ml}(L,M),A_{nk}(L,M)],
\eean
thus
\[
[A_{ml}(L,M),A_{nk}(L,M)]=\sum_{pq}C_{ml,nk}^{pq}A_{pq}(L,M).
\]
This implies that
\[
[\hat{\pa}_{ml},\hat{\pa}_{nk}]W=-[A_{nk}(L,M),A_{ml}(L,M)]_-W
=\sum_{pq}C_{nk,ml}^{pq}\hat{\pa}_{pq}W.\quad \blacksquare
\]
 Noticing that $A_{0,2k}(L,M)=0$ and $A_{0,2k+1}(L,M)=2L^{2k+1}$. Even though
 $\hat{\pa}_{0,2k}L=\hat{\pa}_{0,2k}M=0$, and $\hat{\pa}_{0,2k+1}L=2[L^{2k+1}_+,L]$, however
$\hat{\pa}_{0,2k+1}M=2[L^{2k+1}_+,M]-2(2k+1)L^{2k}$.
Hence $\hat{t}_{0,2l+1} (l\geq 0)$ can not be identified with $t_{2l+1}$ due to the fact that
these additional symmetries are explicitly time-dependent.

Let us define another generator $Y_B(\la,\mu)$ of additional symmetries as\cite{L95}
\bea
Y_B(\la,\mu)&=&\sum_{m=0}^\infty\frac{(\mu-\la)^m}{m!}
\sum_{l=-\infty}^\infty\la^{-l-m-1}(A_{m,m+l}(L,M))_-,
\label{Yml}\\
&=&\sum_{m=0}^\infty\frac{(\mu-\la)^m}{m!}
\sum_{l=-\infty}^\infty\la^{-l-m-1}(M^mL^{m+l}-(-1)^{m+l}L^{m+l-1}M^mL)_-.\no
\eea
We mention that for $\la=\mu$, the generator
$Y_B(\la,\la)=2\sum_{l\in {\mathbb Z\/}_{odd}}\la^{-l-1}L^l_-$
can be viewed as the resolvent operator of the BKP hierarchy.
The main result in this section is the following.
\begin{theorem}
The generator $Y_B(\la,\mu)$ can be expressed as
\[
Y_B(\la,\mu)=\frac{1}{\la}[w(t,-\la)\pa^{-1}\cdot w_x(t,\mu)-w(t,\mu)\pa^{-1}
\cdot w_x(t,-\la)].
\]
\end{theorem}
Proof. Using Lemma \ref{PQ} and the identity
$(P)_-=\sum_{i=1}^{\infty}\pa^{-i}\res_\pa(\pa^{i-1}P)$ for  a pseudodifferential operator
$P=\sum_ia_i\pa^i$ (in particular, for $P=f\pa^{-1}$,
$f\pa^{-1}=\sum_{i=1}^\infty\pa^{-i}f^{(i-1)}$), we have
\bean
(M^mL^{m+l})_-
&=&\sum_{i=1}^\infty\pa^{-i}\res_z[z^{-1}(\pa^{i-1}(M^mW\pa^{m+l+1}e^\xi))((W^*)^{-1}e^{-\xi})],\\
&=&\sum_{i=1}^\infty\res_z[z^{m+l}\pa^{-i}(M^mw)^{(i-1)}\cdot w^*(t,z)],\\
&=&\res_z[z^{m+l}(\pa_z^mw)\pa^{-1}\cdot w^*(t,z)].
\eean
Similarly,
\[
(L^{m+l-1}M^mL)_-=\res_z[z(\pa_z^mz^{m+l-1}w(t,z))\pa^{-1}\cdot w^*(t,z)].
\]
Then
\bean
Y_B(\la,\mu)&=&\sum_{m=0}^\infty\frac{(\mu-\la)^m}{m!}\sum_{l=-\infty}^\infty
\{\la^{-l-m-1}\res_z[z^{m+l}(\pa_z^mw(t,z))\pa^{-1}\cdot w^*(t,z)]\\
&&\quad +(-\la)^{-m-l-1}\res_z[z(\pa_z^mz^{m+l-1}w(t,z))\pa^{-1}\cdot w^*(t,z)]\},\\
&=&\res_z\left[\sum_{n=-\infty}^{\infty}\frac{z^n}{\la^{n+1}}
\sum_{m=0}^\infty\frac{(\mu-\la)^m}{m!}(\pa_z^mw(t,z))\pa^{-1}\cdot w^*(t,z)\right]\\
&&\quad +\frac{\mu}{\la}\res_z\left[\sum_{n=-\infty}^{\infty}\frac{(z+\mu-\la)^n}
{(-\la)^{n+1}}w(t,z+\mu-\la)\pa^{-1}\cdot w^*(t,z)\right],\\
&=&w(t,\mu)\pa^{-1}\cdot w^*(t,\la)+\frac{\mu}{\la}w(t,-\la)\pa^{-1}\cdot w^*(t,-\mu),\\
&=&\frac{1}{\la}[w(t,-\la)\pa^{-1}\cdot w_x(t,\mu)-w(t,\mu)\pa^{-1}\cdot w_x(t,-\la)],
\eean
where we have used the formula
$\res_z(\sum_{n=-\infty}^{\infty}(z^n/\la^{n+1})f(z))=f(\la)$ and (\ref{adjwf}). $\blacksquare$

Based on Dickey's observation \cite{Dic95a,Dic95b}, one can define eigenfunctions $\phi_i(t)$ as
\[
\phi_1(t)=\int d\la\rho_1(\la)w(t,-\la)/\la,\quad
\phi_2(t)=\int d\mu\rho_2(\mu)w(t,\mu),
\]
where $\rho_i(\la)$ are some weighting functions, so that
\[
\pa_{2n+1}\phi_i(t)=B_{2n+1}\phi_i(t),\quad i=1,2.
\]
This enables one to introduce the so-called constrained BKP (cBKP) hierarchy defined
by the Lax operator \cite{YC92,LW99}
\be
K=L^n=L^n_++\phi_1(t)\pa^{-1}\cdot \phi_{2x}(t)-\phi_2(t)\pa^{-1}\cdot \phi_{1x}(t).
\label{laxcBKP}
\ee
It is quite interesting to contrast two kind of reductions of the BKP hierarchy.
For the $n$th-reduced BKP hierarchy ($n=1,3,5,\cdots$)\cite{DJKM,L96}, the associated
Lax operator $L^n=L^n_+$ (or $L^n_-=0$) which is related to inner symmetries of the hierarchy,
while for the cBKP, the negative part $L^n_-$ is constructed from a particular linear combination
of $Y_B(\la,\mu)$ and thus related to additional symmetries of the hierarchy.
It is not hard to verify that the negative part of $K$ in (\ref{laxcBKP})
is indeed compatible with the Lax equation
\[
\pa_{2k+1}K=[K^{\frac{2k+1}{n}}_+,K].
\]
Hence the above discussions provide an origin for the special form of the Lax
operator (\ref{laxcBKP}) from symmetry point of view.

%%%%%%%%%%%%%%%%%%%%%%%%%%%%%%%%%%%%%%%%%%%%%%%%%%%%%%%%%%%%%%%%%%%
\section{Fay identity and Adler-Shiota-van Moerbeke formula}
%%%%%%%%%%%%%%%%%%%%%%%%%%%%%%%%%%%%%%%%%%%%%%%%%%%%%%%%%%%%%%%%%%%%
Having discussed the B\"acklund symmetry and additional symmetries of the BKP hierarchy,
now we like to show that these two symmetries are essentially the same  and are connected by
a kind of ASV formula that was previously proved by van der Leur \cite{L95}.
Our proof relies on the Fay identity and its differential form of $\tau$-functions.

\begin{proposition} \label{Fay}(Fay identity) The $\tau$-function of the BKP hierarchy
satisfies the Fay  quadrisecant  identity:
\bea
&&\sum_{(s_1,s_2,s_3)}\frac{(s_1-s_0)(s_1+s_2)(s_1+s_3)}{(s_1+s_0)(s_1-s_2)(s_1-s_3)}
\tau(t+2[s_0]+2[s_1])\tau(t+2[s_2]+2[s_3])\no\\
&&+\frac{(s_0-s_1)(s_0-s_2)(s_0-s_3)}{(s_0+s_1)(s_0+s_2)(s_0+s_3)}
\tau(t+2[s_0]+2[s_1]+2[s_2]+2[s_3])\tau(t)=0
\label{Fayeq}
\eea
where $(s_1,s_2,s_3)$ stands for cyclic permutations of $s_1$, $s_2$ and $s_3$.
\end{proposition}
Proof. From the bilinear identity (\ref{bileq}) we have
\[
\res_z(z^{-1}G(z)\tau(t)G'(z)\tau(t')e^{\xi(t-t',z)})=\tau(t)\tau(t').
\]
Setting $t=x-y$ and $t'=x+y$ to rewrite it as
\[
\res_z(z^{-1}\tau(x-y-2[z^{-1}])\tau(x+y+2[z^{-1}])e^{-2\xi(y,z)})=\tau(x-y)\tau(x+y).
\]
Now replace $x$ as $x+[s_0]+[s_1]+[s_2]+[s_3]$ and $y$ as $[s_0]-[s_1]-[s_2]-[s_3]$ we have
\bean
&&\res_z\left(z^{-1}\frac{(1-s_0z)(1+s_1z)(1+s_2z)(1+s_3z)}{(1+s_0z)(1-s_1z)(1-s_2z)(1-s_3z)}\right.\\
&&\left.\cdot\frac{\tau(x+2[s_1]+2[s_2]+2[s_3]-2[z^{-1}])}{\tau(x+2[s_1]+2[s_2]+2[s_3])}
\frac{\tau(x+2[s_0]+2[z^{-1}])}{\tau(x+2[s_0])}\right)=1.
\eean
Computing the  residue with poles at $z=\infty, -s_0^{-1}, s_1^{-1}, s_2^{-1}, s_3^{-1}$,
we reach the Fay identity. $\blacksquare$

Just as the case of the KP hierarchy whose quasi-periodic solutions satisfy the Fay
trisecant identity on Jacobian varieties\cite{Kri77,S86}, the Fay quadrisecant identity for the
BKP hierarchy can be viewed as an identity of theta functions on Prym
varieties\cite{S89,Ta97,Kri06}. Next let us derive its differential form.
\begin{proposition} \label{dFay}(Differential Fay identity) The following equation holds.
\bea
&&\left(\frac{1}{s_2^2}-\frac{1}{s_1^2}\right)
\{\tau(t+2[s_1])\tau(t+2[s_2])-\tau(t+2[s_1]+2[s_2])\tau(t)\}\no\\
&&=\left(\frac{1}{s_2}+\frac{1}{s_1}\right)
\{\pa\tau(t+2[s_2])\tau(t+2[s_1])-\pa\tau(t+2[s_1])\tau(t+2[s_2])\}\no\\
&&\quad+\left(\frac{1}{s_2}-\frac{1}{s_1}\right)\{\tau(t+2[s_1]+2[s_2])\pa\tau(t)-
\pa\tau(t+2[s_1]+2[s_2])\tau(t)\}.
\label{dFayeq}
\eea
\end{proposition}
Proof. Setting $s_0=0$ in the Fay identity (\ref{Fayeq}) we have
\bea
&&\tau(t+2[s_1]+2[s_2]+2[s_3])\tau(t)\no\\
&&=\sum_{(s_1,s_2,s_3)}\left(\frac{s_1+s_2}{s_1-s_2}\right)\left(\frac{s_1+s_3}{s_1-s_3}\right)
\tau(t+2[s_2]+2[s_3]])\tau(t+2[s_1])
\label{predFayeq}
\eea
The differential Fay identity (\ref{dFayeq}) can be proved by
differentiating the above equation with respect to $s_3$ and then setting $s_3=0$. $\blacksquare$\\
We remark that the differential Fay identity (\ref{dFayeq}) was also derived by
Takasaki \cite{T06} by differentiating the bilinear identity over $t'_1$ and
setting  $t'=t+2[\la^{-1}]+2[\mu^{-1}]$.

Now we can give a simple proof of the ASV formula for the BKP hierarchy.
\begin{theorem} (Adler-Shiota-van Moerbeke and van de Leur\cite{L95})  The following formula
\be
X_B(\la,\mu)w(t,z)=2\la\left(\frac{\la-\mu}{\la+\mu}\right)Y_B(\la,\mu)w(t,z),
\label{ASV}
\ee
holds, where it should be understood that the vertex operator $X_B(\la,\mu)$
acting on $w(t,z)$ is generated by its action on the $\tau$ function.
\end{theorem}
Proof. The left hand side of (\ref{ASV}) is given by
\bea
&&X_B(\la,\mu)w(t,z)=e^{\xi(t,z)}\left[\frac{\tau(t)G(z)X_B(\la,\mu)\tau(t)-
G(z)\tau(t)\cdot X_B(\la,\mu)\tau(t)}{\tau(t)^2}\right]\no\\
&&=e^{\xi(t,z)-\xi(t,\la)+\xi(t,\mu)}\{\left(\frac{z+\la}{z-\la}\right)
\left(\frac{z-\mu}{z+\mu}\right)\tau(t)\tau(t+2[\la^{-1}]-2[z^{-1}]-2[\mu^{-1}])\no\\
&&\quad-\tau(t-2[z^{-1}])\tau(t+2[\la^{-1}]-2[\mu^{-1}])\}/\tau^2(t),\no\\
&&=e^{\xi(t,z)-\xi(t,\la)+\xi(t,\mu)}\left(\frac{\la-\mu}{\la+\mu}\right)
\{\left(\frac{z+\la}{z-\la}\right)\tau(t-2[\mu^{-1}])\tau(t+2[\la^{-1}]-2[z^{-1}])\no\\
&&\quad -\left(\frac{z-\mu}{z+\mu}\right)\tau(t-2[z^{-1}]-2[\mu^{-1}])
\tau(t+2[\la^{-1}])\}/\tau^2(t),
\label{Xw1}
\eea
where we have use (\ref{predFayeq}) for $s_1=-z^{-1}, s_2=\la^{-1}, s_3=-\mu^{-1}$
to reach the last equality. On the other hand, the right hand side of (\ref{ASV})
is given by
\bea
&&2\la\left(\frac{\la-\mu}{\la+\mu}\right)Y_B(\la,\mu)w(t,z)\no\\
&&=2\left(\frac{\la-\mu}{\la+\mu}\right)\left[w(t,-\la)\pa^{-1}w_x(t,\mu)w(t,z)-
w(t,\mu)\pa^{-1}w_x(t,-\la)w(t,z)\right],\no\\
&&=\left(\frac{\la-\mu}{\la+\mu}\right)\left[w(t,-\la)\pa^{-1}(w_x(t,\mu)w(t,z)-w(t,\mu)w_x(t,z))\right.\no\\
&&\quad\left.-w(t,\mu)\pa^{-1}(w_x(t,-\la)w(t,z)-w(t,-\la)w_x(t,z))\right]
\label{Yw}
\eea
where we have used integration by part to reach the second equality.
Comparing (\ref{Yw}) with (\ref{Xw1}), it is sufficient to prove the following
\bea
&&\left(\frac{z+\la}{z-\la}\right)\left(e^{\xi(t,z)-\xi(t,\la)}
\frac{\tau(t+2[\la^{-1}]-2[z^{-1}])}{\tau(t)}\right)_x\no\\
&&=w(t,-\la)w_x(t,z)-w_x(t,-\la)w(t,z),
\label{asm1}
\eea
and
\bea
&&\left(\frac{z-\mu}{z+\mu}\right)\left(e^{\xi(t,z)+\xi(t,\mu)}
\frac{\tau(t-2[z^{-1}]-2[\mu^{-1}])}{\tau(t)}\right)_x\no\\
&&=w(t,\mu)w_x(t,z)-w_x(t,\mu)w(t,z).
\label{asm2}
\eea
In terms of $\tau$-function, (\ref{asm1}) can be expressed as
\bean
&&(z^2-\la^2)\{\tau(t+2[\la^{-1}]-2[z^{-1}])\tau(t)-\tau(t+2[\la^{-1}])\tau(t-2[z^{-1}])\}\no\\
&&=(z+\la)\{\tau(t+2[\la^{-1}]-2[z^{-1}])\pa\tau(t)-\pa\tau(t+2[\la^{-1}]-2[z^{-1}])\tau(t)\}\no\\
&&\quad +(z-\la)\{\tau(t+2[\la^{-1}])\pa\tau(t-2[z^{-1}])-\pa\tau(t+2[\la^{-1}])\tau(t-2[z^{-1}])\},
\eean
which, after changing of variable $t=t'+2[z^{-1}]$, is just the differential Fay identity
(\ref{dFayeq}) with $s_1=\la^{-1}$ and  $s_2=z^{-1}$. Similarly, (\ref{asm2}) can be verified
in the same manner. $\blacksquare$
\begin{corollary}
The vector fields $\hat{\pa}_{m,m+l}$ of additional symmetries acting on $\tau$-function can be
expressed as
\be
\hat{\pa}_{m,m+l}\tau=Z_l^{(m+1)}(\tau).
\label{asvtau}
\ee
where $Z_l^{(m)}$ are generators of the $W^B_{1+\infty}$-algebra defined by (\ref{Zml}).
\end{corollary}
Proof. Observing the fact that
\[
X_B(\la,\mu)w(t,z)=w(t,z)(G(z)-1)\left[\frac{(X_B(\la,\mu)-1)\tau(t)}{\tau(t)}\right]
\]
and
\[
(A_{m,m+l})_-w(t,z)=-w(t,z)(G(z)-1)\left[\frac{\hat{\pa}_{m,m+l}\tau(t)}{\tau(t)}\right].
\]
Then (\ref{asvtau}) is an immediate consequence of (\ref{Zml}), (\ref{Yml}) and
the ASV formula (\ref{ASV}). $\blacksquare$

Therefore, the $w_{\infty}^B$-algebra  of additional symmetries, acting on Sato operator
(or wave function) is lifted to its central extension, the $W_{1+\infty}^B$-algebra
of B\"acklund symmetries, acting on $\tau$-function.

%%%%%%%%%%%%%%%%%%%%%%%%%%%%%%%%%%%%%%%%%%%%%%%%%%%%%%%%%%%%%%%%%
{\bf Acknowledgments\/}\\
I would like to thank Y.T.Chen for useful discussions.
This work is partially supported by the National Science Council of Taiwan
under Grant No. 95-2112-M-194-005-MY2.
%%%%%%%%%%%%%%%%%%%%%%%%%%%%%%%%%%%%%%%%%%%%%%%%%%%%%%%%%%%%%%%%%%%
%%%%%%%%%%%%%%%%%%%%%%%%%%%%%%%%%%%%%%%%%%%%%%%%%%%%%%%%%%%%%%

\end{document}